\documentstyle[12pt,epsfig]{article}

\def\beq{\begin{equation}}
\def\eeq#1{\label{#1}\end{equation}}
\def\eeqn{\end{equation}}
\def\beqa{\begin{eqnarray}}
\def\eeqa#1{\label{#1}\end{eqnarray}}
\def\eeqan{\end{eqnarray}}
\def\CR{\nonumber \\ }
\def\leqn#1{(\ref{#1})}

\def\stacksymbols #1#2#3#4{\def\theguybelow{#2}
    \def\vp{\lower#3pt}
    \def\sp{\baselineskip0pt\lineskip#4pt}
    \mathrel{\mathpalette\intermediary#1}}

\def\intermediary#1#2{\vp\vbox{\sp
     \everycr={}\tabskip0pt
     \halign{$\mathsurround0pt#1\hfil##\hfil$\crcr#2\crcr
              \theguybelow\crcr}}}

\def\gapproxeq{\stacksymbols{>}{\sim}{2.5}{.2}}
\def\lapproxeq{\stacksymbols{<}{\sim}{2.5}{.2}}

\def\Title#1{\begin{center} {\bf \Huge #1 } \end{center}}
\def\Author#1{\begin{center}{ \sc #1} \end{center}}
\def\Address#1{\begin{center}{ \it #1} \end{center}}
\newenvironment{Abstract}{\begin{quotation} \begin{center}
                       ABSTRACT
     \end{center}\bigskip  }{\end{quotation}}
\def\Acknowledgements{\bigskip  \bigskip \begin{center} \begin{large}
             \bf ACKNOWLEDGEMENTS \end{large}\end{center}}

\def\gp{g^\prime}
\def\hT{\hat{T}}
\def\hS{\hat{S}}
\def\bbox{\stacksymbols{-}{\Box}{1.0}{.2}}

\begin{document}
\begin{titlepage}

\vfill
\Title{Gauge-Assisted Technicolor?}
\vskip2cm
\Author{Maxim Perelstein}
\Address{Institute for High-Energy Phenomenology, Laboratory for Elementary 
Particle Physics, Cornell University, Ithaca, NY, 14853}

\vskip3cm
\begin{Abstract}
It is well known that technicolor models in which the electroweak symmetry is 
broken by QCD-like strong dynamics at the TeV scale generally predict 
unacceptably large corrections to low-energy observables. We investigate the 
models of electroweak symmetry breaking by strong dynamics in which the gauge 
symmetry is extended to include an arbitrary number of additional SU(2) and 
U(1) factors. This class of models includes the deconstructed version of the 
recently proposed five-dimensional "Higgsless" scenario. We conclude that the 
additional structure present in these theories does not suppress the effects 
of strongly coupled short-distance physics on the precision electroweak 
observables. In particular, the possibility that the symmetry breaking is 
due to QCD-like dynamics is still strongly disfavored by data. 
\end{Abstract}
\vfill
\end{titlepage}

\section{Introduction}

Most of the structure of the Standard Model (SM) of electroweak interactions has been experimentally confirmed in the last three decades. One piece of the puzzle, however, is still missing: despite many attempts, the Higgs boson has not been discovered. It 
may well be that the Higgs is simply too heavy for present accelerators, and will be observed once the Large Hadron Collider (LHC) at CERN becomes operational. On the other hand, it is well known that the Higgs hypothesis is plagued with theoretical shortcomings such as the gauge hierarchy problem. In view of this, it is useful to examine whether electroweak physics can be described consistently by models which do not contain a Higgs field (or any other fundamental spin-0 field) at all. 

According to a well-known theorem~\cite{Cornwall}, a theory with massive $W$ and $Z$ gauge bosons and no Higgs field is guaranteed to become strongly coupled at a certain energy scale. In the simplest realization of the theory, which will be reviewed in Section 2, this scale is close to 1 TeV. The full non-perturbative theory is expected to possess additional composite states with masses close to this scale, analogous to heavy mesons such as the $\rho$ and baryons of QCD. These states would contribute to low-energy observables through their interactions with electroweak gauge bosons. While these contributions are not perturbatively calculable, they can be parametrized by including higher-derivative operators in the low-energy theory~\cite{Appel}. Precise agreement of multiple low-energy observables ("precision electroweak observables") with the SM predictions implies constraints on the coefficients of these operators~\cite{Bagger,BPRS}. At the same time, these coefficients can be estimated within the "technicolor" models in which the dynamics at the TeV scale is assumed to be a scaled-up version of QCD~\cite{TC}. The present experimental constraints, notably the constraint from the $S$ parameter~\cite{PT}, strongly disfavor such models.

Given the phenomenological difficulties of the technicolor scenario, is it still possible to construct a viable theory of electroweak symmetry breaking (EWSB) without a Higgs? One possibility is that the strong dynamics at the TeV scale cannot be described by scaling up the familiar strong interactions of QCD, and does not produce a large $S$ parameter. While this is certainly possible, no compelling examples of such strongly coupled systems are known. A possible alternative is a scenario in which the {\it weakly coupled} physics below the cutoff is modified in a way that reduces the effect of the strongly coupled short-distance physics on precision electroweak observables. Such a scenario could potentially revive the attractive possibility of EWSB by QCD-like strong dynamics. As an example, consider the "Higgsless" models~\cite{C1,C2,C3,Yasunori} in which electroweak gauge bosons are allowed to propagate in an extra compact dimension of space, and EWSB occurs at the boundary of this dimension. In these models, the violation of perturbative unitarity in the longitudinal $W$ scattering, often used as an indicator of the onset of strong coupling~\cite{Quigg}, is postponed until an energy scale parametrically above 1 TeV due to cancellations between the contributions of the Kaluza-Klein (KK) gauge bosons. If the scale of strong coupling is indeed raised, one might expect that the low-energy effects of the operators induced at that scale should be suppressed.
The goal of this paper is to determine whether such a suppression actually takes place.

Instead of dealing with the five-dimenional Higgsless models directly, we will study a class of closely related four-dimensional models~\cite{deconHL,Chiv,bess,Evans}. These models possess an extended gauge structure which can in some sense approximate the fifth dimension; this approximation becomes exact in a certain limit~\cite{decon}. On the other hand, these models can be analyzed using the four-dimenisional non-linear sigma model (NLSM) language~\cite{Appel}, making it easier to compare and contrast them with the traditional, four-dimensional theories of higgsless EWSB such as 
technicolor\footnote{A connection between the five-dimensional Higgsless 
models and technicolor, motivated by holography, has been proposed, for 
example, in Ref.~\cite{Barb1}. This connection becomes more direct and explicit
in the four-dimensional models considered here. Precision electroweak 
constraints on the Higgsless models in five dimensions have been obtained in 
Ref.~\cite{Barb1,HLpew}.}. 
The NLSM language allows for clear separation between the perturbatively calculable contributions to the low energy observables from the new weakly-coupled gauge states and the intrinsically uncalculable contributions from the strongly coupled sector. As in the traditional technicolor analyses, the latter can be parametrized by adding higher-derivative operators to the low-energy theory, whose coefficients are constrained by data. Using the method developed recently by Barbieri et. al.~\cite{BPRS}, we will show that, barring unnatural cancellations between low-energy and high-energy contributions, the constraints on the coefficients contributing to the $S$ parameter in models with additional gauge groups are essentially identical to those obtained in four-dimensional models with the minimal gauge structure. In other words, the additional structure {\it does not} serve to suppress the low-energy effects of the strongly coupled sector, and the EWSB by QCD-like strong dynamics is still ruled out. 

\section{Electroweak Symmetry Breaking by Strong Interactions}

\subsection{The Minimal Model}

To set the stage for the analysis of models with extended gauge structure, let us first review the most minimal model of electroweak symmetry breaking without a Higgs boson in four dimensions. In this model, the spectrum of states below the ultraviolet cutoff is identical to the SM without a Higgs. The physics below the cutoff is described by a non-linear sigma model based on an $SU(2)\times SU(2)$ global symmetry group~\cite{Appel}. An $SU(2)\times U(1)$ subgroup of this group is promoted to a local symmetry and is identified with the electroweak gauge symmetry of the SM. At low energies, the lagrangian of the model has the form
\beq
{\cal L} = -\frac{1}{2g^2}\, {\rm Tr}\,W_{\mu\nu}W^{\mu\nu} 
               -\frac{1}{4{\gp}^2} B_{\mu\nu}B^{\mu\nu} + 
               f^2 \,{\rm Tr}\,\left| D_\mu \Sigma \right|^2\,.
\eeq{lagr}
Here the covariant derivative is defined as
\beq
D_\mu \Sigma = \partial_\mu \Sigma + i W_\mu \Sigma - 
i B_\mu \Sigma \tau^3, 
\eeq{der}
where $\tau^a$ are the Pauli matrices with the normalization ${\rm Tr}\,\tau^a\tau^b=\frac{1}{2}\delta^{ab}$; $W_\mu\equiv W_\mu^a\tau^a$ and $B_\mu$ are the $SU(2)_L$ and $U(1)_Y$ gauge fields, respectively (note that we use a non-canonical normalization for these fields); $W_{\mu\nu}$ and $B_{\mu\nu}$ are the corresponding field tensors, and $g$ and $g^\prime$ are the SM $SU(2)_L$ and $U(1)_Y$ coupling constants. The bifundamental scalar field $\Sigma$ acquires a vacuum expectation value (vev), breaking $SU(2)\times SU(2)\to SU(2)$. We choose the normalization so that $\left<\Sigma\right>=$diag$(1,1)$. With this normalization, the charged gauge boson mass is $M_W=gf$, implying $f\approx 124$ GeV. 

 The NLSM is an effective field theory, valid below the cutoff scale $\Lambda\sim4\pi f$. The full NLSM lagrangian should include all operators consistent with the symmetries of the low-energy theory; these operators parametrize the effects of unknown, strongly-coupled physics at energy scales above $\Lambda$, and their coefficients in the lagrangian depend on the details of that physics.
It is convenient~\cite{BPRS} to classify the possible operators according to their symmetry properties with respect to the "custodial" group $SU(2)_c$~\cite{custod} (i.e., the global $SU(2)$ left unbroken by the vev of $\Sigma$), and with respect to the gauged $SU(2)$ identified with the SM weak interaction group $SU(2)_L$.
There are three classes of operators: those breaking both $SU(2)_c$ and $SU(2)_L$ ("class A"); those breaking $SU(2)_L$ but not $SU(2)_c$ ("class B"); and those preserving both symmetries ("class C"). To study the experimental constraints on the theory, it is sufficient to concentrate on the leading (in terms of the derivative expansion) operator in each class.

The only allowed dimension-2 operator (apart from the kinetic term for the $\Sigma$ field included in Eq.~\leqn{lagr}) is in class A; it has the form\footnote{Note that while ${\cal O}_A$ has the same dimension as the kinetic term for $\Sigma$, it is expected to be subdominant in the lagrangian since its coefficient must disappear in the limit where the $U(1)$ gauge field decouples, $\gp\to 0$.} 
 \beq
{\cal O}_A \,=\, \left( {\rm Tr}\, T\Sigma^\dagger D_\mu \Sigma\right)^2,   
 \eeq{Aop}
 where $T=\Sigma\tau_3 \Sigma^\dagger$. The leading class B operator has mass dimension 4:  
 \beq
 {\cal O}_B \,=\, B_{\mu\nu} {\rm Tr} \, T W^{\mu\nu}.
 \eeq{Bop}
The leading class C operators have mass dimension 6:
\beq
{\cal O}_{c,1}\,=\,(\partial_\rho B_{\mu\nu})^2,~~~~
{\cal O}_{c,2}\,=\,2\,{\rm Tr}\,(D_\rho W_{\mu\nu})^2.
 \eeq{cop}
 Retaining only these four operators, the corrections to the NLSM lagrangian can be written as
 \beq 
 \Delta{\cal L} \,=\, f^2 c_A {\cal O}_A \,+\, c_B {\cal O}_b \,+\, \frac{1}{f^2} c_{c,1} {\cal O}_{c,1} \,+\, \frac{1}{f^2} c_{c,2} {\cal O}_{c,2},
\eeq{NLSM_sl} 
where the coefficients $c_i$ are dimensionless.

\subsection{Precision Electroweak Constraints}

The properties of $W/Z$ bosons predicted by the leading-order lagrangian~\leqn{lagr} at tree level are identical to those in the SM with a weakly-coupled Higgs boson. (The leading loop corrections in the two models also match if the NLSM cutoff $\Lambda$ is identified with the Higgs mass.) These properties would be modified, however, in the presence of the additional operators in Eq.~\leqn{NLSM_sl}. Most significant constraints come from the modifications of the $W$ and $Z$ propagators, often referred to as "oblique" corrections. To study these corrections, it is sufficient to only keep the terms in the lagrangian of quadratic order in the gauge fields. These have the form
\beqa
{\cal L} + \Delta {\cal L} &=& \frac{1}{2}\, \left[f^2 - \frac{1}{g^2} q^2 + \frac{4}{f^2} c_{c,2} (q^2)^2 \right] W^a_\mu W^{a\mu} \,+\,\frac{f^2}{4} c_A\,W^3_\mu W^{3\mu} \CR & & + 
\frac{1}{2}\,\left[f^2(1+\frac{c_A}{2}) - \frac{1}{{\gp}^2} q^2 +
\frac{4}{f^2} c_{c,1} (q^2)^2 \right] B^\mu B_\mu \CR & & +\left[-f^2(1+\frac{c_A}{2})\,+c_B\,q^2\right]\,B^\mu W^3_\mu,
\eeqa{quad}
where $W(q)$, $B(q)$ are the Fourier transforms of the corresponding gauge fields. This equation can be used to relate the coefficients $c_i$ to the adimenional form factors introduced in Ref.~\cite{BPRS}:
\beqa
\hS &\equiv& g^2 \Pi^\prime_{W_3B} (0) = - g^2 c_B, \CR
\hT &\equiv& g^2 M_W^{-2} \left(\Pi_{W_3W_3}(0)-\Pi_{W^+W^-}(0)\right) = -\frac{c_A}{2}, \CR
Y &\equiv& \frac{1}{2}{\gp}^2 M_W^2 \Pi_{BB}^{\prime\prime}(0) = -4g^2 {\gp}^2 c_{c,1}, \CR W &\equiv& \frac{1}{2}g^2 M_W^2 \Pi_{W_3W_3}^{\prime\prime}(0) = -4g^4 \,c_{c,2}.
\eeqa{translate}
Here, $\Pi_{ij}(q^2)$ are the elements of the inverse propagator matrix in the basis $(W^\pm, W^3, B)$, and primes denote derivatives with respect to $q^2$. Experimental bounds on the form factors from a global fit to electroweak precision observables, including LEP2 data, were derived in Ref.~\cite{BPRS}. According to Eq.~\leqn{translate}, they translate into  
\beqa
c_A &=& (-2.8\pm2.0)\cdot 10^{-3};
~~~c_B \,=\, (4.0 \pm 3.0)\cdot 10^{-3}; \CR
c_{c,1} &=& (2.3 \pm 5.5) \cdot 10^{-3};~~~
c_{c,2} \,=\, (-0.4 \pm 1.1) \cdot 10^{-3}.  
\eeqa{bounds}
These bounds were obtained using the value of the cutoff scale $\Lambda=800$ GeV; the bounds are only logarithmically sensitive to this value.

\subsection{Technicolor Models}

What are the implications of these bounds for specific models with EWSB by a  strongly coupled sector? Answering this question would require computing the coefficients $c_A, c_B, c_{c,1}, c_{c,2}$ in specific models. In general, such calculations are not yet feasible. An important exception is provided by the technicolor models, in which the dynamics at the TeV scale is a rescaled version of the familiar strong interactions of QCD. The $S$ parameter in these models was estimated by Peskin and Takeuchi~\cite{PT} using dispersion relations, large-$N$ techniques, and the experimental data on processes such as $e^+e^-\to \pi^+\pi^-, \tau\to\nu_\tau \pi^+ 2\pi^-$, etc. For a model based on an $SU(N_{\rm TC})$ gauge group with $N_{\rm TF}$ flavors of technifermions ($q_L$ and $q_R$ count as two separate flavors), these authors obtain\footnote{The $S$ parameter used in Ref.~\cite{PT} is related to $\hS$ by a simple rescaling, $S=4s_W^2 \hS/\alpha \approx 119 \hS$.}
\beq
\hS \approx (2.5\cdot 10^{-3})\,\frac{N_{\rm TF}}{2}\,\frac{N_{\rm TC}}{3}.
\eeq{TCshat}
According to Eq.~\leqn{translate}, this implies that the operator ${\cal O}_B$ in these models is induced with a coefficient
\beq
c_B\,\approx\,(-6\cdot 10^{-3})\,\frac{N_{\rm TF}}{2}\,\frac{N_{\rm TC}}{3}.
\eeq{TCcoeff}
Comparing this result with Eq.~\leqn{bounds} shows that even the most minimal technicolor model, $N_{\rm TC}=3$, $N_{\rm TF}=2$,  
is ruled out at more than 3 $\sigma$.

Strictly speaking, Eq.~\leqn{TCcoeff} should be interpreted as the value of $c_B$ at the scale $M_Z$, since the dispersion relations include the perturbative contributions which renormalize this coefficent between $M_Z$ and $\Lambda$. At the same time, the bounds in Eq.~\leqn{bounds} apply to the values of the coefficients at $\Lambda$. The above discussion ignores this difference; it can be taken into account by subtracting the one-loop perturbative renormalization contribution to $\hS$, $\hS_{\rm pert}\approx\frac{\alpha}{24\pi s_W^2}\log(\Lambda/M_Z) \approx 1.0\cdot 10^{-3}$, from Eq.~\leqn{TCshat}. Including this correction does not change the qualitative conclusions reached above. More generally, it is clear that any significant cancellation between the perturbative and the non-preturbative  (UV) contributions would imply fine tuning. Since the goal of the models with extended gauge structure is precisely to avoid fine tuning, we will ignore the loop-level perturbative contributions to $\hS$ in the analysis of those models. The obtained numerical bounds will be on the same footing as Eq.~\leqn{TCcoeff}: while not exact, they will nevertheless provide the level of accuracy sufficient for our purposes.

\section{A Prototype Model with an Extended Gauge Sector: $SU(2)\times SU(2)\times U(1)$}

The model reviewed above is minimal in the sense that it contains no states below the UV cutoff beyond those of the SM without a Higgs. We would like to extend this model in a way that could potentially reduce the effects of the operators in Eq.~\leqn{NLSM_sl} on the low-energy observables. One possibility that appears promising is to extend the gauge sector of the model to $SU(2)^{n+1}\times U(1)_Y$. The extended group can be broken down to $U(1)_{\rm em}$ by strong dynamics, without the need for Higgs fields. In the limit $n\to\infty$, these models become identical to the five-dimensional models of~\cite{C1,C2}, provided that the gauge coupling constants and symmetry breaking parameters are chosen appropriately. In the 5D models, the breakdown of perturbative unitarity in $W_L$ scattering occurs at a higher scale than in the minimal four-dimensional model of Section 2 due to cancellations involving KK modes of the $W/Z$ bosons. Similar cancellations should occur in the 4D $SU(2)^{n+1}\times U(1)$ model, with the extra massive gauge bosons playing the role of the KK states. It may be expected that raising the strong coupling scale reduces the effects of the physics at that scale on low-energy observables. Our goal is to investigate whether this is in fact the case.

\begin{figure}[t]
\begin{center}
\epsfig{file=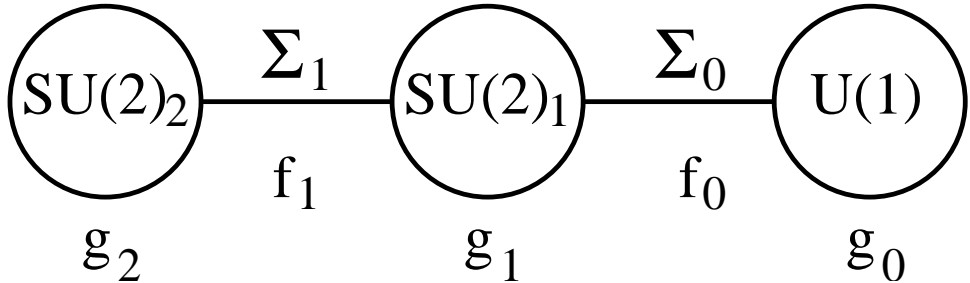}
\caption{The diagram representing the symmetry breaking pattern of the model studied in Section 3. Each circle indicates a global $SU(2)$ symmetry; the gauged subgroup is indicated inside the circle. The lines connecting the circles represent the bifundamental scalar fields $\Sigma_{0,1}$ whose vacuum expectation values break the symmetries.}
\label{fig:221}
\end{center}
\end{figure}

Before presenting the analysis for arbitrary $n$, let us discuss the simplest case, $n=1$, in some detail. The main points of this discussion will carry over to the more general case. Consider an NLSM based on the $SU(2)^3\to SU(2)$ symmetry breaking pattern; the surviving $SU(2)$ plays the role of the custodial symmetry. The
gauged subgroup is $SU(2)^2\times U(1)$. The symmetry breaking is achieved by two bifundamental $\Sigma$ fields, and can be conveniently depicted by the diagram in Fig.~\ref{fig:221}. The leading-order lagrangian has the form
\beq
{\cal L} = -\sum_{i=1}^2 \frac{1}{2g_i^2}\, {\rm Tr}\,W^{(i)}_{\mu\nu}W^{(i)\mu\nu} 
               -\frac{1}{4g_0^2} B_{\mu\nu}B^{\mu\nu} + 
               f_0^2 \,{\rm Tr}\,\left| D_\mu \Sigma_0 \right|^2 +
               f_1^2 \,{\rm Tr}\,\left| D_\mu \Sigma_1 \right|^2,
\eeq{lagr_ext1} 
where the covariant derivatives are defined as
\beqa
D_\mu \Sigma_0 &=& \partial_\mu \Sigma_0 + i W^{(1)}_\mu \Sigma_0 - i B_\mu \Sigma_0 \tau^3, \CR
D_\mu \Sigma_1 &=& \partial_\mu \Sigma_1 + i W^{(2)}_\mu \Sigma_1 - i \Sigma_1 W^{(1)}_\mu .
\eeqa{der_ext1}
Here $W^{(i)}\equiv W^{(i)a}\tau^a$ and $g_i$ $(i=1,2)$ are the gauge fields and the coupling constants, respectively, of the two gauged $SU(2)$ factors, and $g_0$ is the $U(1)$ coupling constant. The $\Sigma$ fields acquire vevs of the form $\left<\Sigma_0\right>=\left<\Sigma_1\right>=$diag$(1,1)$, resulting in charged gauge boson masses
\beq
M^2_{\pm} \,=\,\frac{1}{2}\left( g_1^2 (f_0^2+f_1^2) + g_2^2 f_1^2 \pm
\sqrt{(g_1^2 (f_0^2+f_1^2) + g_2^2 f_1^2)^2 - 4g_1^2g_2^2f_0^2f_1^2} \right).
\eeq{masses}
The SM $W$ boson is identified with the lighter of these two states.
The neutral gauge boson sector contains a massless photon and two massive states, the lighter of which is identified with the $Z$.

Unlike the minimal model of Section 2, the leading-order lagrangian of Eq.~\leqn{lagr_ext1} {\it does not} predict $W/Z$ properties identical to the SM, due to mixing between these states and the additional, heavier gauge bosons. The effects of this mixing on precision electroweak observables have been investigated in Ref.~\cite{Chiv,bess}. In addition to this, however, $W/Z$ properties are sensitive to the strongly coupled physics beyond the regime of validity of the NLSM, which should be parametrized by adding higher-derivative operators to the lagrangian in analogy with Eq.~\leqn{NLSM_sl}. Again, the operators can be classified according to their symmetry properties. Given the importance of the $S$ parameter in constraining technicolor models, we will concentrate on the operators that respect the custodial $SU(2)_c$, but break some of the gauge symmetries. At the leading (dimension-4) level, there are three such operators:
\beqa
{\cal O}_1 &=& B^{\mu\nu} \,{\rm Tr}\, (W^{(1)}_{\mu\nu}\Sigma_0\tau^3\Sigma_0^\dagger),~~~
{\cal O}_2 \,=\, {\rm Tr}\,(W^{(1)}_{\mu\nu}\Sigma_1^\dagger W^{(2)\mu\nu} \Sigma_1), \CR & &
{\cal O}_3 \,=\, B^{\mu\nu} \,{\rm Tr}\, (W^{(2)}_{\mu\nu}\Sigma_1\Sigma_0\tau^3\Sigma_0^\dagger\Sigma_1^\dagger).
\eeqa{opers1} 
We will consider corrections to the lagrangian~\leqn{lagr_ext1} of the form
\beq
\Delta {\cal L} \,=\, c_1 {\cal O}_1 \,+\, c_2 {\cal O}_2 \,+\, c_3{\cal O}_3.
\eeq{NLSM_sl1}
While not guaranteed, it is quite reasonable to expect that $c_3$ is much smaller than the other two coefficients. Indeed, within perturbation theory ${\cal O}_3$ is not generated until the three-loop level, requiring both sets of pions to enter the diagram, while both ${\cal O}_1$ and ${\cal O}_2$ are generated at one loop. We will set 
$c_3=0$ for the remainder of the analysis.

 An efficient way of computing the corrections to precision electroweak observables in the model defined by Eqs.~\leqn{lagr_ext1},~\leqn{NLSM_sl1} is to apply the method proposed by Barbieri~et.~al. in Ref.~\cite{BPRS}. The model under consideration belongs to the class of "universal" models defined by these authors. The left-handed SM fermions have to transform as doublets under one of the gauge groups, $SU(2)_1$ or $SU(2)_2$, and singlets under the other group. For concreteness, let us first consider the case when they transform under $SU(2)_2$, with the lagrangian
\beq
{\cal L}_{\rm int} \,=\,\bar{\Psi}\gamma^\mu\,(W^{(2)}_\mu + Y B_\mu)\Psi.
\eeq{int}
While $W^{(2)}$ is not a mass eigenstate, it is convenient to choose it as an interpolating field for the light gauge bosons. Integrating out $W^{(1)}$ (i.e., replacing this field with the solution of its equation of motion), one obtains the inverse propagator matrix in the basis $(W^{(2)\pm}, W^{(2)3}, B)$, from which the form factors $\hS, \hT$ etc. can be easily found. As in Section 2, it is sufficient to work at the quadratic level in the gauge fields, with the lagrangian of the form
\beqa
{\cal L} + \Delta {\cal L} &=& -\sum_{i=1}^2 \frac{1}{2g_i^2} q^2  W^{(i)a}_\mu W^{(i)a\mu} - \frac{1}{2 g_0^2} q^2 B_\mu B^\mu \CR & &+\, \frac{f_0^2}{2} \left[ 2 W_\mu^{(1)\pm}
W^{(1)\mp\,\mu} + (W_\mu^{(1)3}-B_\mu)(W^{(1)3\,\mu}-B^\mu) \right]
\CR & & +\,\frac{f_1^2}{2} (W_\mu^{(2)a}-W_\mu^{(1)a})(W^{(2)a\,\mu}-W^{(1)a\,\mu}) \CR & & + \,c_1 q^2 B^\mu W_\mu^{(1)3} \,+\, c_2 q^2 W^{(1)a\,\mu}W_\mu^{(2)a}.
\eeqa{quad1}
The equations of motion for $W^{(1)}$ have the form
\beqa
\frac{1}{g_1^2}\,q^2\,W^{(1)\pm}_\mu &=& f_0^2\,W^{(1)\pm}_\mu \,-\,
f_1^2\,(W^{(2)\pm}_\mu-W^{(1)\pm}_\mu)\,+\, c_2 q^2 W^{(2)\pm}_\mu,\CR
\frac{1}{g_1^2}\,q^2\,W^{(1)3}_\mu &=&  f_0^2\,(W^{(1)3}_\mu-B_\mu) -f_1^2\,(W^{(2)3}_\mu-W^{(1)3}_\mu)\CR & &+ q^2 (c_1B_\mu + c_2 W^{(2)3}_\mu).
\eeqa{W1eom}
Solving these equations for $W^{(1)\pm}, W^{(1)3}$ and substituting the results into~\leqn{quad1} yields
\beqa
{\cal L} + \Delta {\cal L} &=& \frac{1}{2}\,\left[ -\frac{q^2}{g_2^2} + \frac{f_1^2(q^2-g_1^2 f_0^2-2c_2g_1^2q^2)}{q^2-g_1^2(f_0^2+f_1^2)} \right]\,W^{a\mu}W_\mu^a \CR
& & +\,\frac{1}{2}\left[ -\frac{q^2}{g_0^2} + \frac{f_0^2(q^2-g_1^2 f_1^2-2c_1 g_1^2 q^2) }{q^2-g_1^2(f_0^2+f_1^2)} \right]\,B^\mu B_\mu \CR & &+\,\frac{g_1^2 (f_0^2 f_1^2-c_1 f_1^2 q^2 - c_2 f_0^2 q^2) }{q^2-g_1^2(f_0^2+f_1^2)}\,W^3_\mu B^\mu,
\eeqa{quad1_1}
where $W\equiv W^{(2)}$ is the interpolating field for the light $W/Z$ bosons, and terms of order $c_i^2$ have been dropped. From this lagrangian one can read off the elements of the inverse propagator matrix $\Pi_{ij}(q^2)$, and therefore the adimensional form factors defined in Eq.~\leqn{translate}. The most interesting constraints come from the $\hS$ parameter. Defining the mixing angle $\theta$ such that $\tan\theta = f_1/f_0$, we obtain 
\beq
\hS \,=\, \frac{g^2}{g_1^2}\,\cos^2 \theta \sin^2\theta \,-\, g^2(c_1\sin^2\theta+c_2\cos^2\theta).
\eeq{hats1}
The two terms in this equation have very different origins: the first term represents the contribution of the additional heavy gauge bosons present in the theory defined by Eq.~\leqn{lagr_ext1}, whereas the second term is generated by the short-distance physics that lies outside the regime of validity of the NLSM. The second term in turn contains two distinct contributions from the operators ${\cal O}_1$ and ${\cal O}_2$. In the absence of fine tuning between the three contributions,  the experimental bound on $\hat{S}$ implies\footnote{While the first term in Eq.~\leqn{hats1} is positive-definite, the $\hS$ parameter is required to be negative at 1 $\sigma$ level. We list a 2 $\sigma$ constraint on $\tan\theta$ in Eq.~\leqn{bound_1}. Note also that the experimental bound on $\hS$ used here 
takes into account the perturbative one-loop contributions computed under the assumption that the SM is valid up to the cutoff at  800 GeV~\cite{BPRS}. In the model considered here, there will be an additional perturbative contribution due to the loops with heavy gauge bosons; we ignore this contribution, as it is unlikely to affect any of our conclusions.}  
\beqa
& & \hskip-1cm \tan\theta \gapproxeq 5.2,~~c_1 \approx (4.0\pm 3.0)\cdot 10^{-3},~~~\frac{c_2}{\tan^2\theta} \approx (4.0\pm 3.0)\cdot 10^{-3};\CR
& &{\rm or}~~~~\CR
& & \hskip-1cm \tan\theta \lapproxeq 0.19,~~c_1 \tan^2\theta \approx (4.0\pm 3.0)\cdot 10^{-3},~~c_2 \approx (4.0\pm 3.0)\cdot 10^{-3},
\eeqa{bound_1}   
where we have used the weak-coupling condition $g_1^2\lapproxeq 4\pi$.
The physical meaning of this result is obvious: the decoupling of the extra gauge bosons requires a hierarchy of the breaking scales, either $f_0 \gg f_1$ ($\theta\approx 0$) or $f_0 \ll f_1$ ($\theta \approx \pi/2$). After integrating out the heavy gauge bosons, the theory reduces to the model of Section 2: in the first case, the light $W$ bosons nearly coincide with the $SU(2)_2$ gauge eigenstate, whereas in the second case they consist predominantly of the diagonal combination of the $SU(2)_1$ and $SU(2)_2$ eigenstates. In each case, one of the dimension-4 operators, ${\cal O}_1$ or ${\cal O}_2$, plays the role of ${\cal O}_B$ of Section 2, and its coefficient is strongly constrained. The coefficient of the other operator is virtually unconstrained; nevertheless, we conclude that the additional gauge structure present in the model does not reduce the overall sensitivity to the dimension-4 operators contributing to the $S$ parameter. 

In the spirit of the technicolor approach, the model considered here can be thought of as arising from a QCD-like theory with a product gauge group, $SU(N_{\rm TC,0})\times SU(N_{\rm TC,1})$, and matter  
in the following representations: $q_0\in (\Box,1)$, $\bar{q}_0\in (\bbox,1)$, $q_1\in (1,\Box)$, $\bar{q}_1\in (1,\bbox)$. Under the global symmetries of the NLSM, $q_0\in (\Box,1,1)$, $\bar{q}_0\in (1,\Box,1)$, $q_1\in (1,\Box,1)$, $\bar{q}_1\in (1,1,\Box)$. The NLSM symmetries are broken by vacuum condensates $\left< q_0 \bar{q}_0 \right>$ and
$\left< q_1 \bar{q}_1\right>$; these are parametrized by $\Sigma_0$ and $\Sigma_1$, respectively. In such a model, one expects the coefficients $c_1$ and $c_2$ to be approximately given by Eq.~\leqn{TCcoeff}, with the appropriate values of $N_{\rm TC}$ and $N_{\rm TF}$. By the same logic as in Section 2, this model appears to be strongly disfavored. Moreover, even allowing for fine tuning between different terms in Eq.~\leqn{hats1} does not improve the situation, since within this model all three terms have the same (positive) sign. 

For completeness, let us list the expressions for the other three adimensional form factors in the model at hand:
\beqa
\hT &=& 0;~~~Y = \,\frac{g^2g^{\prime 2}}{g_1^4}\,\cos^4\theta \sin^2\theta \left( \cos^2\theta - 2 g_1^2 c_1 \right);\CR
W &=& \,\frac{g^4}{g_1^4}\,\cos^2\theta \sin^4\theta \left( \sin^2\theta - 2 g_1^2 c_2 \right).
\eeqa{others}
The $\hT$ parameter vanishes due to the unbroken custodial symmetry of the model. The experimental constraints from $W$ and $Y$ parameters are parametrically weaker than the bounds in Eq.~\leqn{bound_1}.

Finally, consider the case when the left-handed SM fermions are charged under $SU(2)_1$ instead of $SU(2)_2$. Integrating out the field $W_2^{(2)}$ yields
\beqa
\hS &=& -g^2 c_1,~~~~ \hT\,=\,Y\,=\,0, \CR
W &=& \,\frac{g^4}{g_2^4}\,\cos^2\theta\,(1-2c_2 g_2^2).
\eeqa{1pars}
Consistency with experimental constraints requires $\cos\theta \ll 1$, implying that the group $SU(2)_2$ has to be broken well above the scale of EWSB ($f_1\gg f_0$). Below the scale $f_1$, the theory reduces to the model of Section 2, with the operator ${\cal O}_1$ playing the role of ${\cal O}_B$. It is not surprising that the coefficient of this operator is constrained at exactly the same level as $c_B$ by the bound on the $\hS$ parameter. Again, the additional gauge structure does not relax the constraint on the short-distance physics, and a technicolor-like realization of this model is ruled out. 

\section{A More General Gauge Structure:  $SU(2)^{n+1}\times U(1)$}

\begin{figure}[t]
\begin{center}
\epsfig{file=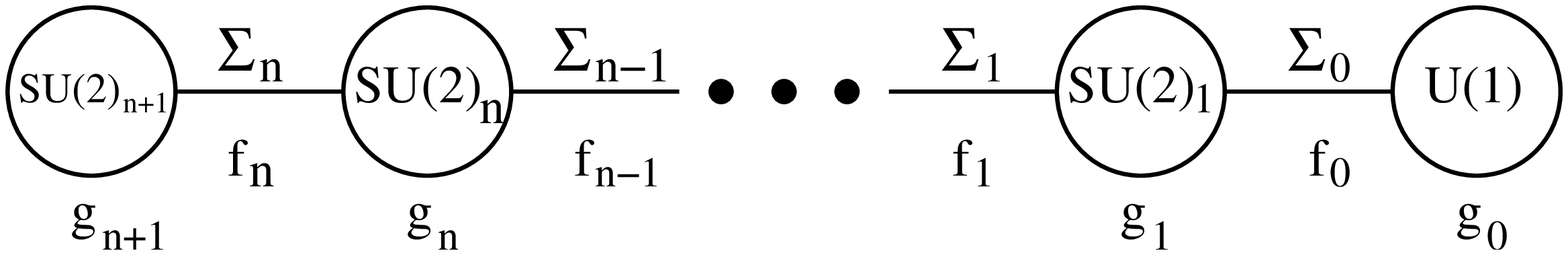,scale=0.6}
\caption{The diagram representing the symmetry breaking pattern of the models studied in Section 4. The notation is the same as in Fig.~\ref{fig:221}.}
\label{fig:n}
\end{center}
\end{figure}

Let us generalize the analysis of the previous section to the case of a larger gauge group, $SU(2)^{n+1}\times U(1)_Y$ with arbitrary $n$. In analogy with the model of Section 3, consider the NLSM describing the $SU(2)^{n+2}\to SU(2)$ global symmetry breaking by a chain of $n+1$ bifundamental fields, as shown in Fig.~\ref{fig:n}. An $SU(2)^{n+1}\times U(1)$ subgroup of the original symmetry is gauged. The leading-order lagrangian of the model has the form
\beq
{\cal L} \,=\, -\sum_{i=1}^{n+1} \frac{1}{2g_i^2}\,{\rm Tr}\,W_{\mu\nu}^{(i)} W^{(i)\mu\nu}\,-\,\frac{1}{4g_0^2}\,B_{\mu\nu} B^{\mu\nu} \,+\, \sum_{k=0}^n f_k^2\,{\rm Tr}\,|D_\mu \Sigma_k|^2,
\eeq{lagr_n}  
where the covariant derivatives are defined as
\beqa
D_\mu \Sigma_0 &=& \partial_\mu \Sigma_0\,+\,iW_\mu^{(1)}\Sigma_0\,-\,iB_\mu \Sigma_0 \tau^3,\CR
D_\mu \Sigma_k &=& \partial_\mu \Sigma_k \,+\,iW_\mu^{(k+1)}\Sigma_k\,-\,i\Sigma_k W_\mu^{(k)},~~~k=1 \ldots n.
\eeqa{ders_n}
The bifundamental fields acquire vevs of the form $\left<\Sigma_j\right>=$diag$(1,1)$, $j=0\ldots n$, leading to $n+1$ massive charged and neutral gauge bosons (the lightest of which are identified with the SM $W/Z$) plus a massless photon.

Once again, the lagrangian of Eq.~\leqn{lagr_n} should be supplemented by adding all other operators consistent with the NLSM symmetries to account for the effects of strongly-coupled, short-distance physics. We will concentrate on the leading (dimension-4) operators that contribute to the $S$ parameter, and involve only the pairs of "neighboring" gauge groups (see the diagram of Fig.~\ref{fig:n}):
\beq
\Delta {\cal L} = \sum_{i=0}^{n} c_{i,i+1} {\cal O}_{i,i+1}
\eeq{corr_n}
with
\beqa
{\cal O}_{0,1} &=& B^{\mu\nu}\,{\rm Tr}\,(W^{(1)}_{\mu\nu}\Sigma_0 \tau^3 \Sigma_0^\dagger), \CR
{\cal O}_{i,i+1} &=& {\rm Tr}\,(W^{(i)}_{\mu\nu}\Sigma_i^\dagger W^{(i+1)\mu\nu} \Sigma_i),~~~i=1\ldots n.
\eeqa{opers_n}
We do not include operators of the same type involving non-neighboring gauge fields, such as, for example, ${\rm Tr}\,(W^{(i-1)}_{\mu\nu}\Sigma_{i-1}^\dagger \Sigma_i^\dagger W^{(i+1)\mu\nu} \Sigma_i \Sigma_{i-1})$. This is justified by the same argument as the omission of ${\cal O}_3$ in Section 3: while not guaranteed, it is likely that the coefficients of such operators are subdominant to those included in~\leqn{corr_n} since in perturbation theory they are not generated at the one-loop level. At the quadratic level in the gauge fields, the lagrangian defined by Eqs.~\leqn{lagr_n},~\leqn{corr_n} reads
\beqa
{\cal L}+\Delta{\cal L} &=& -\,\sum_{i=1}^{n+1}\frac{1}{2g_i^2}\,q^2 W_\mu^{(i)a} W^{(i)a\mu}\,-\frac{1}{2g_0}\,q^2 B_\mu B^\mu\CR
& &+\,\frac{f_0^2}{2}\left[ 2 W_\mu^{(1)\pm}
W^{(1)\mp\,\mu} + (W_\mu^{(1)3}-B_\mu)(W^{(1)3\,\mu}-B^\mu) \right]
\CR & & +\,\sum_{j=1}^n \frac{f_j^2}{2} (W_\mu^{(j+1)a}-W_\mu^{(j)a})(W^{(j+1)a\,\mu}-W^{(j)a\,\mu}) \CR & & + 
\,c_{0,1} q^2 B^\mu W_\mu^{(1)3} \,+\, \sum_{k=1}^{n} c_{k,k+1} q^2 W^{(k)a\,\mu}W_\mu^{(k+1)a}.
\eeqa{quad_n} 
The left-handed fermions of the SM are charged under the $U(1)$ and one of the $SU(2)$ factors; let us first consider the case when this factor is $SU(2)_{n+1}$, the leftmost group in Fig.~\ref{fig:n}. We would like to integrate out the gauge fields $W^{(i)}_\mu$ with $i=1
\ldots n$; we will use an iterative approach to obtain formulae valid for arbitrary $n$. First, integrate out $W^{(n)}_\mu$. The equation of motion for this field is
\beqa
\frac{1}{g_n^2}\,q^2\,W^{(n)a}_\mu &=& -f_n^2 \left( W_\mu^{(n+1)a}-W_\mu^{(n)a} \right) \,+\,f_{n-1}^2 \left( W_\mu^{(n)a}-W_\mu^{(n-1)a} \right) \CR & &+\,q^2 \left( c_{n,n+1}W^{(n+1)a}_\mu + c_{n-1,n} W^{(n-1)a}_\mu \right),
\eeqa{eom_nn} 
with the solution
\beq
W^{(n)a}_\mu \,=\, G(q^2) \left[ (f_n^2 -c_{n,n+1}q^2)\,W^{(n+1)a}_\mu \,+\,(f_{n-1}^2 - c_{n-1,n} q^2)\,W^{(n-1)a}_\mu\right],
\eeq{sol_nn}
where
\beq
G(q^2) \,=\, \frac{-g_n^2}{q^2-g_n^2(f_n^2+f_{n-1}^2)}.
\eeq{g_func}
In the analysis of this section, we will be only interested in extracting the $\hS$ parameter. (Note that $\hT=0$ due to custodial symmetry.) Since  $\hS\propto \Pi^\prime_{W_3B}(0)$, it will suffice to only keep the terms up to the linear order in $q^2$. Plugging the solution~\leqn{sol_nn} back into Eq.~\leqn{quad_n} and dropping the terms of order $(q^2)^2$ and higher yields
\beqa
{\cal L}+\Delta{\cal L} &=& -\sum_{i=1}^{n-2} \frac{1}{2g_i^2}\,q^2 W^{(i)2} \,-\,\frac{1}{2\tilde{g}_{n-1}^2}
\,q^2 W^{(n-1)2} - \frac{1}{2\tilde{g}_{n+1}^2}
\,q^2 W^{(n+1)2}\CR & &-\frac{1}{2g_0^2}\,q^2 B^2 +
\frac{f_0^2}{2}\left( W^{(1)2} -2 B W^{(1)3} + B^2 \right) \CR 
& &+\sum_{j=1}^{n-2} \frac{f_j^2}{2} \left(W^{(j+1)}-W^{(j)}\right)^2 +\,
\frac{\tilde{f}_{n-1}^2}{2}\,\left(W^{(n+1)}-W^{(n-1)}\right)^2 \CR 
& &+c_{0,1} q^2 B W^{(1)3} + \sum_{k=1}^{n-2} c_{k,k+1} q^2 W^{(k)}W^{(k+1)} \CR & &+\tilde{c}_{n-1,n+1} q^2 W^{(n-1)} W^{(n+1)},
\eeqa{quad_n1} 
where the sums over the Lorentz indices and, in terms not involving $B$, the color indices are implicit: $W^{(i)2}\equiv W^{(i)a}_\mu W^{(i)a\mu}$, $B^2\equiv B_\mu B^\mu$, etc. We have defined
\beqa
\frac{1}{\tilde{g}_{n-1}^2} &=& \frac{1}{g_{n-1}^2} \,+\,\frac{\cos^4\theta_n}{g_n^2} - 2c_{n-1,n}\cos^2 \theta_n; \CR
\frac{1}{\tilde{g}_{n+1}^2} &=& \frac{1}{g_{n+1}^2} \,+\, \frac{\sin^4\theta_n}{g_n^2} - 2 c_{n,n+1} \sin^2\theta_n;\CR
\tilde{f}_{n-1}^2 &=& \frac{f_{n-1}^2 f_n^2}{f_{n-1}^2+f_n^2};\CR
\tilde{c}_{n-1,n+1} &=& -\frac{1}{g_n^2}\,\sin^2\theta_n \cos^2\theta_n
+ c_{n-1,n}\,\sin^2\theta_n + c_{n,n+1}\,\cos^2\theta_n,
\eeqa{formulas}
where $\tan\theta_n=f_n/f_{n-1}$.

The iterative approach is based on the observation that Eq.~\leqn{quad_n1} has {\it exactly} the same structure as the original lagrangian, Eq.~\leqn{quad_n}. This should not be surprising, since Eq.~\leqn{quad_n} already contains all the terms (up to order $q^2$) that are compatible with the symmetries of the model and are "local" in the sense that only interactions between the gauge fields adjacent to each other in the diagram of Fig.~\ref{fig:n} are allowed. Eliminating the field $W^{(n)}$ results only in the renormalization of the gauge couplings $g_{n-1}$ and $g_{n+1}$, and the appearance of two new terms (proportional to $\tilde{f}_{n-1}$ and $\tilde{c}_{n-1,n+1}$) that describe the interactions between the new neighbors, $W^{(n-1)}$ and $W^{(n+1)}$. The structure of these interactions is identical to those between any other pair of neighboring gauge fields. 

Starting with Eq.~\leqn{quad_n1} and integrating out successively the fields $W^{(n-1)}$, $W^{(n-2)}$, $\ldots$, $W^{(2)}$ leads to a lagrangian of the form
\beqa
{\cal L}+\Delta{\cal L} &=& - \frac{1}{2\tilde{g}_1^2}\,q^2 W^{(1)2} \,-\, \frac{1}{2\tilde{g}_{n+1}^2}
\,q^2 W^{(n+1)2} -\frac{1}{2g_0^2}\,q^2 B^2 \CR & &+
\frac{f_0^2}{2}\left( W^{(1)2} -2 B W^{(1)3} + B^2 \right) +
\frac{\tilde{f}_1^2}{2}\,\left(W^{(n+1)}-W^{(1)}\right)^2 \CR 
& &+c_{0,1} q^2 B W^{(1)3} +\tilde{c}_{1,n+1} q^2 W^{(1)} W^{(n+1)},
\eeqa{lagr_n2}
which is identical (up to notational changes) to the model studied in Section 3 (see Eq.~\leqn{quad1}).
The parameters $\tilde{g}_1$, $\tilde{g}_{n+1}$, $\tilde{f}_1$, and
$\tilde{c}_{1,n+1}$ can be expressed recursively in terms of the parameters entering the original lagrangian, Eq.~\leqn{quad_n}. In particular, we obtain
\beq
\tilde{c}_{1,n+1}\,=\,\sum_{k=2}^n \left( -\frac{1}{\tilde{g}_k^2} + \frac{c_{k-1,k}}{\cos^2\theta_k} \right) \sin^2\theta_k \prod_{j=2}^k \cos^2\theta_j \,+\,c_{n,n+1} \prod_{j=2}^n \cos^2\theta_j,
\eeq{c_coef}
where 
\beq
\frac{1}{\tilde{g}_k^2}\,=\, \frac{1}{g_k^2}\,+\,\sum_{j=k+1}^n \left(
\frac{1}{g_j^2} - \frac{2c_{j-1,j}}{\cos^2\theta_j}\right) \prod_{l=k+1}^j
\cos^4\theta_l,~~~~k=1\ldots n-1,
\eeq{gtildes}
and $\tilde{g}_n\equiv g_n$. The mixing angles $\theta_k$ are defined by $\tan\theta_k = \tilde{f}_k/f_{k-1}$, $k=2\ldots n$, where $\tilde{f}_k^2/2$ is the coefficient of the term $(W^{(k)}-W^{(n+1)})^2$ in the intermediate lagrangian at the stage of the iteration process when the $W^{(k)}$ field is integrated out. We will not need explicit formulae for $\tilde{f}_k$ and $\tilde{g}_{n+1}$.

The remaining step of integrating out the field $W^{(1)}$ and extracting $\hS$ has already been performed in Section 3; in analogy with Eq.~\leqn{hats1}, we obtain
\beq
\hS = \frac{g^2}{\tilde{g}_1^2}\,\cos^2\theta_1 \sin^2\theta_1 \,-\, g^2\left(c_{0,1}\sin^2\theta_1+\tilde{c}_{1,n+1} \cos^2\theta_1\right),
\eeq{hS_n}
where $\tan\theta_1=\tilde{f}_1/f_0$. Finally, combining Eqs.~\leqn{c_coef}
and~\leqn{hS_n} yields
\beq
\hS \,=\, g^2\left[\sum_{k=1}^n \left( \frac{1}{\tilde{g}_k^2} - \frac{c_{k-1,k}}{\cos^2\theta_k}\right) \sin^2\theta_k \, \prod_{j=1}^{k} \cos^2\theta_j \,-\,c_{n,n+1} \prod_{j=1}^n \cos^2\theta_j \right].
\eeq{final}
As in the analysis of Section 3, there are two groups of terms in Eq.~\leqn{final}: those due to the mixing of the weakly coupled heavy gauge bosons with $W$ and $Z$, and those due to the strongly coupled short-distance physics parametrized by the operators in Eq.~\leqn{corr_n}. Again, we require that terms of each group {\it separately} satisfy the experimental constraint\footnote{The comments in footnote 4 apply in this case as well.},
\beq
\hS \lapproxeq 0.9\times 10^{-3}.
\eeq{hS2sigma}
In other words, we assume that there is no finely tuned cancellation between the two groups. The mixing terms have the form
\beq
\hS^{(1)} \,=\, g^2\,\sum_{k=1}^n \left[ \frac{1}{g_k^2}+ \sum_{j=k+1}^n 
\frac{1}{g_j^2} \prod_{l=k+1}^j \cos^4\theta_l \right] \sin^2\theta_k \, \prod_{j=1}^{k} \cos^2\theta_j.
\eeq{group1}
Since each term in the sum of Eq.~\leqn{group1} is positive, any subset of terms must also satisfy the bound~\leqn{hS2sigma}. In particular, the constraints
\beq
\sum_{k=1}^m \sin^2\theta_k \prod_{j=1}^k \cos^2\theta_j \lapproxeq 0.03, ~~~m=1\ldots n,
\eeq{con_gb} 
must be satisfied. (Here we used the weak coupling condition, $g_k^2\lapproxeq 4\pi$.) It can be shown (see Appendix A) that these constraints imply that either $\cos^2 \theta_l \lapproxeq 0.03$ for at least one value of $l\in[1,n]$, or $\sin^2 \theta_l\lapproxeq 0.03$ for all $l$. Physically, this means that a certain amount of hierarchy between the scale of EWSB and the scale at which the additional symmetries are broken is required to suppress the mixing effects.   

The short-distance contribution to $\hS$ has the form
\beq
 \hS^{(2)} \,=\, -g^2 \sum_{k=1}^{n+1} \beta_{k} c_{k-1,k},
\eeq{bdef}
where the coefficients $\beta_k$ can be read off from Eqs.~\leqn{final},~\leqn{gtildes}. Assuming no fine tuning between different operators, we obtain a constraint
\beq
\beta_k c_{k-1,k} \,=\, (4.0 \pm 3.0)\times 10^{-3},~~~k=1 \ldots n+1.
\eeq{constr}
This equation clearly shows that the sensitivity of $\hS$ to short-distance physics in the model considered here would be reduced, compared to the minimal model of Section 2, if and only if the coefficients $\beta_k$ are {\it small} for all $k$. At the same time, using Eqs.~\leqn{final} and ~\leqn{gtildes}, one can show that these coefficients obey the following inequalities:
\beqa
\beta_k &\geq& \tan^2 \theta_k \prod_{j=1}^k \cos^2 \theta_j,~~~k=1 \ldots n; \CR
\beta_{n+1} &\geq& \prod_{j=1}^n \cos^2 \theta_j.
\eeqa{ineq_beta}
It is shown in the Appendix A that these inequalities, together with the constraints~\leqn{con_gb}, imply that 
\beq
\beta_l\gapproxeq 0.97
\eeq{main}
for {\it at least one} value of $l\in[1,n+1]$. Thus, at least one of the operators in Eq.~\leqn{corr_n} gives an unsuppressed contribution to the $\hS$ parameter. 

Consider a technicolor-like ultraviolet completion of the theory discussed here that generalizes the model described in Section 3. 
The completion is a gauge theory based on a product group $\prod_{i=0}^{n} SU(N_{{\rm TC},i})$, with $N_{{\rm TF},i}$ flavors of techniquarks $q_i, \bar{q}_i$ in fundamental and antifundamental representations of the $i$th gauge factor. The $\Sigma$ fields of the model parametrize the techniquark pair operators acquiring vevs at low energies, $\Sigma_i \sim \left< q_i \bar{q}_i \right>$. In this theory, one expects that the operators listed in Eq.~\leqn{opers_n} will be generated with coefficients approximately given by Eq.~\leqn{TCcoeff} with the appropriate values of $N_{\rm TC}$ and $N_{\rm TF}$:
\beq
c_{i,i+1}\,\approx\,(-6\cdot 10^{-3})\,\frac{N_{{\rm TF},i}}{2}\,\frac{N_{{\rm TC},i}}{3}.\eeq{TClike}
Since at least one of these coefficients is constrained by data as strongly as $c_B$ in the minimal model of Section 2, this class of models is as strongly disfavored as the minimal model. (In fact, using the fact that all the coefficients $c_{i,i+1}$ are generated with the same sign and the inequality $\sum_{k=1}^{n+1} \beta_k \geq 1$, which follows from Eq.~\leqn{ineq_beta}, one can show that {\it each} of the coefficients has to satisfy the bound identical to that on $c_B$.) Thus, the additional gauge structure present in the model does not revive the possibility of a technicolor-like ultraviolet completion. This conclusion is valid even if one allows for fine-tuning between the short-distance and mixing contributions to $\hS$: within this class of models, these two contributions have the same sign and no cancellation is possible.

The above analysis assumed that the left-handed SM fermions are charged under $SU(2)_{n+1}$, the leftmost gauge group in the "chain" (see Fig.~\ref{fig:n}). The generalization to the case when they are instead charged under one of the "intermediate" gauge groups, $SU(2)_m$, is straightforward. Notice that starting with the lagrangian~\leqn{quad_n} and integrating out the field $W^{(n+1)}$ yields 
\beqa
{\cal L}+\Delta{\cal L} &=& -\,\sum_{i=1}^{n-1}\frac{1}{2g_i^2}\,q^2 W_\mu^{(i)a} W^{(i)a\mu}\,-\frac{1}{2\tilde{g}_n^2}\,q^2 W_\mu^{(n)a} W^{(n)a\mu}\,-\,
\frac{1}{2g_0}\,q^2 B_\mu B^\mu\CR
& &+\,\frac{f_0^2}{2}\left[ 2 W_\mu^{(1)\pm}
W^{(1)\mp\,\mu} + (W_\mu^{(1)3}-B_\mu)(W^{(1)3\,\mu}-B^\mu) \right]
\CR & & +\,\sum_{j=1}^{n-1} \frac{f_j^2}{2} (W_\mu^{(j+1)a}-W_\mu^{(j)a})(W^{(j+1)a\,\mu}-W^{(j)a\,\mu}) \CR & & + 
\,c_{0,1} q^2 B^\mu W_\mu^{(1)3} \,+\, \sum_{k=1}^{n-1} c_{k,k+1} q^2 W^{(k)a\,\mu}W_\mu^{(k+1)a},
\eeqa{quad_mod}
where 
\beq
\frac{1}{\tilde{g}_n^2}\,=\,\frac{1}{g_{n+1}^2}+\frac{1}{g_n^2}-2c_{n,n+1}.
\eeq{renorm}
This lagrangian has a structure identical to~\leqn{quad_n}, and the procedure of integrating out the leftmost field (now $W^{(n)}$) can be repeated. Proceeding in this way, one can show that integrating out the fields $W^{(m+1)}, \ldots, W^{(n+1)}$ results in a lagrangian of the form~\leqn{quad_n}, with the replacement $n\to m$ and $g_m \to
\tilde{g}_m$ given by a straightforward generalization of Eq.~\leqn{renorm}. Since the SM fermions are now charged under the "leftmost" remaining group, $SU(2)_m$, the problem reduces to the case considered  above. Therefore, the conclusions of the above analysis, such as the bound~\leqn{main}, apply in these cases as well.  

\begin{figure}[t]
\begin{center}
\epsfig{file=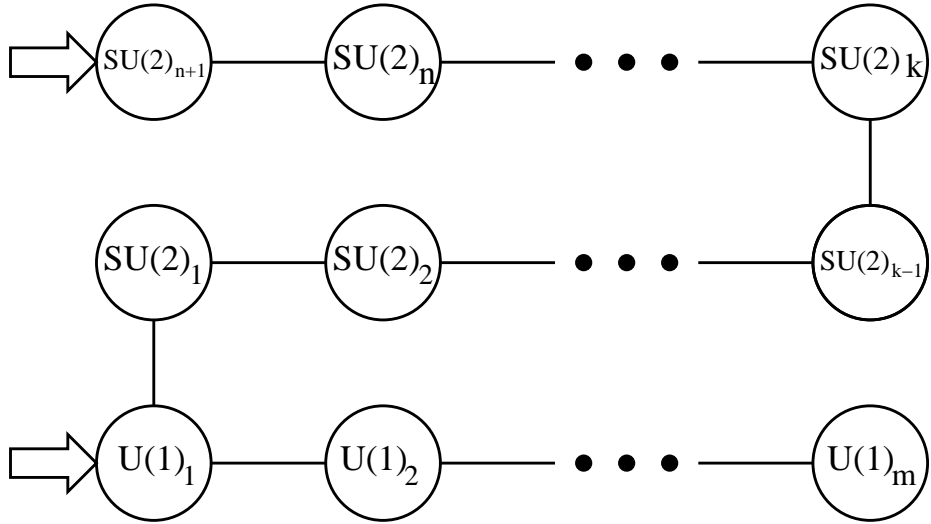,scale=0.6}
\caption{The diagram representing the symmetry breaking pattern of the $SU(2)^{n+1}\times U(1)^m$ model. The diagram is drawn in a way suggested by the interpretation of this model as the deconstructed version of a five-dimensional $SU(2)\times SU(2) \times U(1)$ gauge theory. The arrows indicate the groups under which the light fermions of the model are charged.}
\label{fig:nm}
\end{center}
\end{figure}

Finally, consider the model~\cite{Chiv} based on an $SU(2)^{n+1}\times U(1)^m$ gauge group with the light fermions coupled to the leftmost $SU(2)$ and $U(1)$ factors (see Fig.~\ref{fig:nm}). This model is a deconstruction of the five-dimensional theory with the bulk $SU(2)\times SU(2)\times U(1)$ gauge group, such as the warped-space model of Ref.~\cite{C2}. The conclusions of the above analysis apply to this model as well. Indeed, integrating out the additional $U(1)$ factors reduces the model to the one considered above, the only modification being that the coupling $g_0$ has to be replaced with its renormalized value that depends on the $U(1)_k$ gauge couplings and the $U(1)\times U(1)\to U(1)$ symmetry breaking scales. This modification does not affect the
subsequent analysis.

\section{Conclusions}

The analysis of this paper shows that extending the gauge structure of the models of EWSB by strong interactions in a way suggested by the five-dimensional "Higgsless" constructions does not reduce the sensitivity of the low-energy observables involving $W/Z$ bosons to the strongly coupled short-distance physics. In other words, the experimental constraints on the coefficients of the operators induced by short-distance physics are as strong as they are in the minimal four-dimensional model with no additional particles below the cutoff. In view of this result, extending the gauge structure of the model appears quite unmotivated. This is the main conclusion of this work.

It should be noted that this negative conclusion can be avoided if one allows for cancelations between the contributions to the $S$ parameter from the weakly coupled extra gauge bosons and the strongly coupled short-distance physics. Such cancellations, however, are impossible if the short-distance physics is QCD-like, since in this case the two contributions have the same sign. Thus, even allowing for fine tuning, the possibility of EWSB by a new QCD-like sector is ruled out.  

The models studied here can be thought of as a "latticized" version of five-dimensional theories, and in the limit $n\to\infty$ they give rise to continuum five-dimensional theories. In this limit, the operators in Eq.~\leqn{opers_n}, which played a crucial role in our analysis, 
correspond to terms of the form $(\partial_\mu \partial_5 {\bf W}_\nu)^2$, where ${\bf W}_N$ is the bulk $SU(2)$ gauge field.   
The corresponding term in the bulk lagrangian has the form
\beq
\Delta {\cal L} \sim \frac{1}{\Lambda^2}~{\rm Tr}\,(D_P {\bf W}_{MN})^2  
\eeq{5Daction}
where $\Lambda$ is the cutoff scale. Our analysis suggests that such terms give substantial contributions to the $S$ parameter in theories with localized 
fermions, regardless of the details of the bulk geometry and even when the 
five-dimensional gauge coupling is position-dependent. This generalizes the 
results of Ref.~\cite{Barb1,HLpew}, where it was found that the strong 
coupling scale $\Lambda$ in five-dimensional warped-space models with 
constant gauge couplings cannot be raised far above TeV while maintaining 
consistency with precision electroweak constraints. More generally, in any 
five-dimensional Higgsless model, the experimental lower bound on the scale 
$\Lambda$ can only be obtained by considering the contribution of the local 
operators such as the one in Eq.~\leqn{5Daction} to precision electroweak 
observables. Including such operators in the analysis is important for 
establishing the viability of such models.

\Acknowledgements

I would like to thank Nima Arkani-Hamed for emphasizing the usefulness of the deconstructionist approach in the analysis of the five-dimensional higgsless theories of EWSB. I am also grateful to Giacomo Cacciapaglia for useful comments. This research is supported by the NSF grant PHY-0355005.

\appendix
\section{Proof of the Inequality~\leqn{main}}

In this Appendix, we prove that the inequalty~\leqn{main} is valid for at least one value of $l\in[1,n+1]$. Consider the constraint in Eq.~\leqn{con_gb} for $m=1$:  
\beq
\sin^2\theta_1 \cos^2\theta_1 \lapproxeq 0.03.
\eeq{con1} 
There are two solutions to this inequality: $\cos^2\theta_1\lapproxeq 0.03$ (case 1, $\theta_1\approx\pi/2$), and $\sin^2\theta_1\lapproxeq 0.03$ (case 2, $\theta_1\approx0$). At the same time, according to Eq.~\leqn{ineq_beta}, $\beta_1\geq\sin^2\theta_1$. In the case 1, $\sin^2\theta_1\gapproxeq 0.97$ and the inequality~\leqn{main} is satisfied for $l=1$, completing the proof. In the case 2, consider the $m=2$ constraint in Eq.~\leqn{con_gb}:
\beq
\delta_1 + (1-\delta_1) \sin^2\theta_2 \cos^2 \theta_2 \lapproxeq 0.03,
\eeq{con2}
where $\delta_1=\sin^2\theta_1$ and terms of order $\delta_1^2$ are neglected. Again, there are two solutions: $\cos^2\theta_2 \lapproxeq 
0.03-\delta_1 \leq 0.03$ (case 2.1, $\theta_2\approx\pi/2$) and $\sin^2 \theta_2 \lapproxeq 0.03 - \delta_1$ (case 2.2, $\theta_2\approx 0$). According to~\leqn{ineq_beta}, $\beta_2\geq \sin^2\theta_2 \cos^2 \theta_1$. In the case 2.1, this implies $\beta_2 \gapproxeq (0.97+\delta_1)(1-\delta_1) = 0.97 + {\cal O}(\delta^2)$, and the inequality~\leqn{main} is satisfied for $l=2$, completing the proof. In the case 2.2, we go on to consider the $m=3$ constraint in Eq.~\leqn{con_gb}, and so on. In general, if the constraints of Eq.~\leqn{con_gb} for $k=1\ldots (m-1)$ are solved by making the corresponding angles small, $\theta_k\approx 0$, then the $k=m$ constraint has to be considered. It is solved either by $\theta_m\approx\pi/2$ (or, more precisely,
$\cos^2\theta_m \lapproxeq 0.03-\sum_{i=1}^{m-1}\delta_m$, where $\delta_m=\sin^2\theta_m$ are small), or by $\theta_m\approx 0$ $(
\sin^2\theta_m \lapproxeq 0.03-\sum_{i=1}^{m-1}\delta_m)$. In the 
first case, $\beta_m$ satisfies~\leqn{main} and the proof is completed:
\beq
\beta_m \geq \sin^2\theta_m \prod_{j=1}^{m-1}\cos^2\theta_j
\gapproxeq \left(0.97+\sum_{i=1}^{m-1}\delta_m \right)\prod_{j=1}^{m-1} (1-\delta_j) = 0.97 + {\cal O}(\delta^2).
\eeq{conm}
In the second case, we repeat the analysis for the $(m+1)$th constraint. If $m=n$ is reached and the inequality~\leqn{main} has not been shown to hold for any $l$, it must hold for $l=n+1$. Indeed, in this case, all the angles $\theta_i$ are small, and $\sum_{i=1}^n \delta_m\lapproxeq 0.03$; but this implies that $\beta_{n+1}\geq
\prod_{j=1}^n \cos^2\theta_j \gapproxeq 0.97$, concluding the proof.


\begin{thebibliography}{99}

\bibitem{Cornwall}
J.~M.~Cornwall, D.~N.~Levin and G.~Tiktopoulos,
Phys.\ Rev.\ D {\bf 10}, 1145 (1974)
[Erratum-ibid.\ D {\bf 11}, 972 (1975)].

\bibitem{Appel}
T.~Appelquist and C.~W.~Bernard,
Phys.\ Rev.\ D {\bf 22}, 200 (1980);
A.~C.~Longhitano,
Phys.\ Rev.\ D {\bf 22}, 1166 (1980);
Nucl.\ Phys.\ B {\bf 188}, 118 (1981).

\bibitem{Bagger}
J.~A.~Bagger, A.~F.~Falk and M.~Swartz,
Phys.\ Rev.\ Lett.\  {\bf 84}, 1385 (2000)
[arXiv:hep-ph/9908327].

\bibitem{BPRS}
R.~Barbieri, A.~Pomarol, R.~Rattazzi and A.~Strumia,
arXiv:hep-ph/0405040.

\bibitem{TC}
S.~Dimopoulos and L.~Susskind,
Nucl.\ Phys.\ B {\bf 155}, 237 (1979);
L.~Susskind,
Phys.\ Rev.\ D {\bf 20}, 2619 (1979);
S.~Weinberg,
Phys.\ Rev.\ D {\bf 19}, 1277 (1979).

\bibitem{PT}
M.~E.~Peskin and T.~Takeuchi,
Phys.\ Rev.\ Lett.\  {\bf 65}, 964 (1990);
Phys.\ Rev.\ D {\bf 46}, 381 (1992).

\bibitem{C1}
C.~Csaki, C.~Grojean, H.~Murayama, L.~Pilo and J.~Terning,
Phys.\ Rev.\ D {\bf 69}, 055006 (2004)
[arXiv:hep-ph/0305237].

\bibitem{C2}
C.~Csaki, C.~Grojean, L.~Pilo and J.~Terning,
Phys.\ Rev.\ Lett.\  {\bf 92}, 101802 (2004)
[arXiv:hep-ph/0308038].

\bibitem{C3}
C.~Csaki, C.~Grojean, J.~Hubisz, Y.~Shirman and J.~Terning,
Phys.\ Rev.\ D {\bf 70}, 015012 (2004)
[arXiv:hep-ph/0310355].

\bibitem{Yasunori}
Y.~Nomura,
JHEP {\bf 0311}, 050 (2003)
[arXiv:hep-ph/0309189].

\bibitem{Quigg}
B.~W.~Lee, C.~Quigg and H.~B.~Thacker,
Phys.\ Rev.\ Lett.\  {\bf 38}, 883 (1977);
Phys.\ Rev.\ D {\bf 16}, 1519 (1977).

\bibitem{deconHL}
R.~Foadi, S.~Gopalakrishna and C.~Schmidt,
JHEP {\bf 0403}, 042 (2004)
[arXiv:hep-ph/0312324].

\bibitem{Chiv}
R.~S.~Chivukula, E.~H.~Simmons, H.~J.~He, M.~Kurachi and M.~Tanabashi,
arXiv:hep-ph/0406077;
R.~S.~Chivukula, H.~J.~He, J.~Howard and E.~H.~Simmons,
Phys.\ Rev.\ D {\bf 69}, 015009 (2004)
[arXiv:hep-ph/0307209].

\bibitem{bess}
R.~Casalbuoni, S.~De Curtis and D.~Dominici,
arXiv:hep-ph/0405188.

\bibitem{Evans}
N.~Evans and P.~Membry,
arXiv:hep-ph/0406285.

\bibitem{decon}
N.~Arkani-Hamed, A.~G.~Cohen and H.~Georgi,
Phys.\ Rev.\ Lett.\  {\bf 86}, 4757 (2001)
[arXiv:hep-th/0104005];
C.~T.~Hill, S.~Pokorski and J.~Wang,
Phys.\ Rev.\ D {\bf 64}, 105005 (2001)
[arXiv:hep-th/0104035].

\bibitem{Barb1}
R.~Barbieri, A.~Pomarol and R.~Rattazzi,
Phys.\ Lett.\ B {\bf 591}, 141 (2004)
[arXiv:hep-ph/0310285].

\bibitem{HLpew}
H.~Davoudiasl, J.~L.~Hewett, B.~Lillie and T.~G.~Rizzo,
Phys.\ Rev.\ D {\bf 70}, 015006 (2004)
[arXiv:hep-ph/0312193];
JHEP {\bf 0405}, 015 (2004)
[arXiv:hep-ph/0403300];
G.~Burdman and Y.~Nomura,
Phys.\ Rev.\ D {\bf 69}, 115013 (2004)
[arXiv:hep-ph/0312247];
G.~Cacciapaglia, C.~Csaki, C.~Grojean and J.~Terning,
arXiv:hep-ph/0401160.

\bibitem{custod}
P.~Sikivie, L.~Susskind, M.~B.~Voloshin and V.~I.~Zakharov,
Nucl.\ Phys.\ B {\bf 173}, 189 (1980).

\end{thebibliography}
\end{document}